\documentclass{ws-ijmpa}

\begin{document}

\markboth{H.-H. Adam et al.} {New results on the
pd$\rightarrow$$^3$He $\eta$ production near threshold}

\title{New results on the pd$\rightarrow$$^3$He $\eta$ production near threshold}

\author{\footnotesize H.-H. ADAM\footnote{Email: adamh@uni-muenster.de}, A. KHOUKAZ, N. LANG, T. LISTER\footnote{Present address: Hermann-Holthusen-Institut, Lohm\"{u}hlenstr. 5, Hamburg, 20099, Germany}, R. SANTO, S.
STELTENKAMP\footnote{Present address: Institut f\"{u}r
Physikalische Chemie, Universit\"{a}t Mainz, Jakob-Welder-Weg
11-15, Mainz, 55128, Germany}}
\address{Institut f\"{u}r Kernphysik, Universit\"{a}t M\"{u}nster, Wilhelm-Klemm-Str. 9\\
M\"{u}nster, 48149, Germany}

\author{\footnotesize R.~CZY{\.Z}YKIEWICZ, M.~JANUSZ, L.~JARCZYK, B.~KAMYS,
P.~MOSKAL, C.~PISKOR--IGNATOWICZ, J.~PRZERWA, J.~SMYRSKI }
\address{Nuclear Physics Department, Jagellonian University,
Cracow, 30-059, Poland}

\author{\footnotesize D.~GRZONKA, K.~KILIAN, W.~OELERT, T.~SEFZICK, P.~WINTER, M.~WOLKE, P.~W{\"U}STNER}
\address{IKP and ZEL Forschungszentrum J{\"u}lich,
J{\"u}lich, 52425, Germany}

\author{\footnotesize A.~BUDZANOWSKI}
\address{Institute of Nuclear Physics,
 Cracow, 31-342, Poland}

\author{\footnotesize T.~RO{\.Z}EK, M.~SIEMASZKO, W.~ZIPPER}
\address{Institute of Physics, University of Silesia,
Katowice, 40-007, Poland}

\maketitle

\pub{Received (Day Month Year)}{Revised (Day Month Year)}

\begin{abstract}
Measurements on the $\eta$ meson production in proton-deuteron
collisions have been performed using the COSY-11 facility at COSY
(J{\"u}lich). Here we present preliminary results on total and
differential cross sections for the pd$\rightarrow$$^3$He $\eta$
reaction at five excess energies between Q = 5.1 and Q = 40.6 MeV.
The obtained angular distributions for the emitted $\eta$ mesons
in the center of mass system expose a transition from an almost
isotropic emission to a highly anisotropic distribution. The
extracted total cross sections support a strong $\eta$-$^3$He
final state interaction and will be compared with model
predictions.

\keywords{near threshold meson production; final state
interaction; quasi-bound states.}
\end{abstract}

\section{Introduction}

Close to threshold data on the pd$\rightarrow$$^{3}$He $\eta$
reaction are of great interest to study the strong attractive
$\eta$-nucleus interaction at low energies, which might be a
signal for the existence of quasi-bound $\eta$-nucleus states.
First observed at the SPES-IV and SPES-II spectrometers at
SATURNE\cite{berger,mayer}, the $\eta$-production in the reaction
channel pd$\rightarrow$$^{3}$He $\eta$ reveals remarkable
features. Additionally to the unexpected high cross section, the
shape of the excitation function reveals a maximum very close to
the production threshold and a large drop of the production
amplitude within only a few MeV above threshold, which clearly
deviates from pure phase space expectations. In contrast, the
angular distributions of the emitted $\eta$ mesons in the center
of mass system appeared to be consistent with pure phase space
expectations and exhibit no contributions from higher partial
waves.

In order to describe this unexpected near threshold behaviour, a
two-step model based on a double-scattering reaction mechanism has
been proposed by Kilian and Nann\cite{kilian}. Calculations by
F{\"a}ldt and Wilkin\cite{wilkin}, exploiting this approach,
succeeded to reproduce the threshold cross section within a factor
of 2.4. However, in order to reproduce the observed rapid drop of
the production amplitude with increasing energy, the two-step
approach had to be refined by a strong $\eta$-$^{3}$He final state
interaction (FSI) with a large $\eta$-$^{3}$He scattering length.

Further measurements performed at higher excess energies of
$\sim$50 MeV by the COSY-GEM-Collaboration\cite{betigeri} and
between 22 MeV and 120 MeV by the WASA/PROMICE
collaboration\cite{bilger}, resulted in non-isotropic angular
distributions and total cross sections that are underestimated by
the description of the refined two-step model fitted to the SPES
data. A different reaction mechanism based on the excitation of
the N*(1535) has been suggested\cite{betigeri}, however, it fails
to reproduce the observed shape of the previously determined
excitation function.

Therefore, to clarify the still open question concerning the
dominant production process, as well as to investigate the
development of the angular distributions with increasing excess
energy, additional measurements on the reaction of type
pd$\rightarrow$$^{3}$He\,X have been carried out using the COSY-11
installation.

\section{Experiment and Results}

The reaction channel pd$\rightarrow$$^{3}$He $\eta$ has been
studied using the COSY-11 installation\cite{brauksiepe} at COSY
(J\"{u}lich)\cite{cosy} by detection of the emitted $^3$He nuclei
and identification and reconstruction of the produced $\eta$
mesons using the missing mass technique. Over the range of studied
excess energies (Q=5.1, 10.8, 15.2, 20.0 and 40.6 MeV) a
transition from a rather flat angular distribution at the two
lowest energies to a highly anisotropic behaviour for the highest
measured excess energy of 40.6 MeV is observed.

\begin{figure}
\centerline{\psfig{file=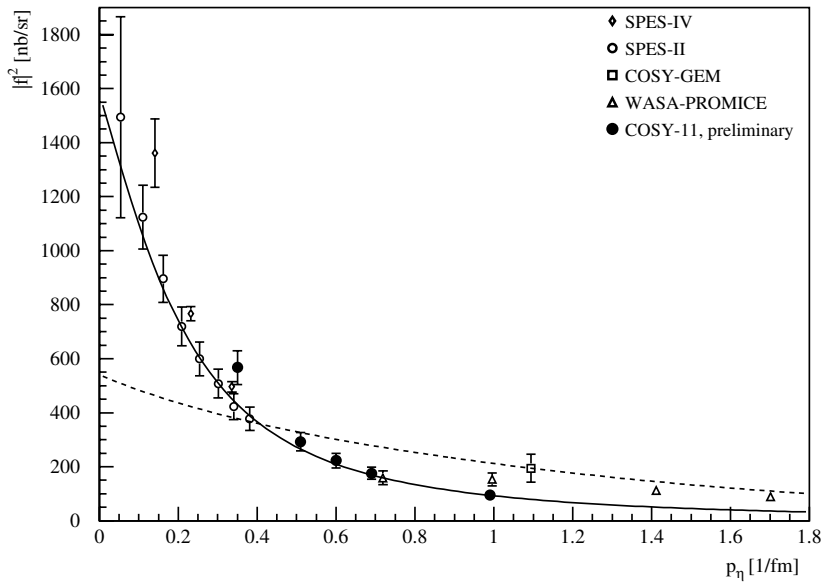,width=10.9 cm}} \vspace*{8pt}
\caption{Average production amplitude squared $\mid$f$\mid^{2}$ as
fuction of the cms momentum of the emitted $\eta$ mesons for the
pd$\rightarrow$$^{3}$He $\eta$ reaction. The solid and dashed
lines represent fits to the data based on a strong $\eta$-$^{3}$He
FSI and a resonance model description respectively. }
\label{amplitude}
\end{figure}

By determining the integrated luminosity via the pd$\rightarrow$pd
elastic scattering\cite{adam,steltenkamp} it was also possible to
derive total cross sections and production amplitude information
for each of the measured excess energies. The obtained production
amplitudes close the open gap between the SATURNE data at low
energies and the higher energy data from WASA/PROMICE and GEM
(figure \ref{amplitude}). The observed shape of the COSY-11 data
strongly supports the predictions of the refined two-step model
and indicates none or only weak contributions from a production
according to the resonance model.

\section*{Acknowledgments}
This work was supported by the FFE grants (41266606 and 41266654)
from the Forschungszentrum J{\"u}lich, the European Community -
Access to Research Infrastructure action of the Improving Human
Potential Programme, by the DAAD Exchange Programme (PPP-Polen)
and the Polish State Committe for Scientific Research (grants No.
2P03B07123 and PB1060/P03/2004/26).


\begin{thebibliography}{0}
\bibitem{berger} J. Berger et al., {\it Phys. Rev. Lett.} {\bf 61}, 919 (1988).
\bibitem{mayer} B. Mayer et al., {\it Phys. Rev.} {\bf C 53}, 2068 (1996).
\bibitem{kilian} K. Kilian and H. Nann, {\it Particles and Fields, AIP Conf. Proc.} {\bf 221}, 185 (1990).
\bibitem{wilkin}G. F\"{a}ldt and C. Wilkin, {\it Phys. Lett.} {\bf B 221}, 20 (1995).
\bibitem{betigeri} M. Betigeri et al., {\it Phys. Lett.} {\bf B 472}, 267 (2000).
\bibitem{bilger} R. Bilger et al., {\it Phys. Rev.} {\bf C 65}, 44608 (2002).
\bibitem{brauksiepe} S. Brauksiepe et al., {\it Nucl. Instr. and Meth.} {\bf A 376}, 397 (1996).
\bibitem{cosy} R. Maier, {\it Nucl. Instr. and Meth.} {\bf A 390}, 1 (1997).
\bibitem{dombrowski} H. Dombrowski et al., {\it Nucl. Phys.} {\bf A 626}, 427c (1997).
\bibitem{adam} H.-H. Adam, Diploma Thesis, Universit\"{a}t M\"{u}nster, Germany, 2000.
\bibitem{steltenkamp} S. Steltenkamp, Diploma Thesis, Universit\"{a}t M\"{u}nster, Germany, 2002.
\end{thebibliography}
\end{document}